\documentclass{article}
\pdfoutput=1
\usepackage{spconf,amsmath,epsfig}
\usepackage{color}
\usepackage{multicol}
\usepackage{multirow}
\usepackage{amssymb}
\usepackage{amsthm}
\usepackage{subfigure}
\usepackage{graphicx}
\usepackage{float} 
\let\OLDthebibliography\thebibliography
\renewcommand\thebibliography[1]{
  \OLDthebibliography{#1}
  \setlength{\parskip}{0pt}
  \setlength{\itemsep}{0pt plus 0.3ex}
}

\begin{document}\sloppy
\def\x{{\mathbf x}}
\def\L{{\cal L}}
\def\etal{\emph{et al.}}
\def\ie{\emph{i.e.}}
\def\eg{\emph{e.g.}}

\newcommand{\jiang}[1]{{\color{cyan}#1}}
\newcommand{\bihan}[1]{{\color{black}#1}}
\newtheorem{theorem}{Theorem}

\title{R3L: Connecting Deep Reinforcement Learning to Recurrent Neural Networks for Image Denoising via Residual Recovery}

\name{Rongkai Zhang$^{1}$, Jiang Zhu$^{1}$, Zhiyuan Zha$^{1}$, Justin Dauwels$^{2}$ and Bihan Wen$^{1 *}$\thanks{$^*$Bihan Wen (bihan.wen@ntu.edu.sg) is the corresponding author.}\thanks{Copyright 2021 IEEE. Published in the IEEE 2021 International Conference on Image Processing (IEEE ICIP 2021), scheduled for 19-22 September 2021 in Anchorage, Alaska, United States. Personal use of this material is permitted. However, permission to reprint/republish this material for advertising or promotional purposes or for creating new collective works for resale or redistribution to servers or lists, or to reuse any copyrighted component of this work in other works, must be obtained from the IEEE. Contact: Manager, Copyrights and Permissions / IEEE Service Center / 445 Hoes Lane / P.O. Box 1331 / Piscataway, NJ 08855-1331, USA. Telephone: + Intl. 908-562-3966.}}
\address{$^{1}$School of Electrical and Electronic Engineering, Nanyang Technological University, Singapore \\ $^{2}$Department of Microelectronics, Delft University of Technology, Netherlands}

\ninept
\maketitle
\begin{abstract}
State-of-the-art image denoisers exploit various types of deep neural networks via deterministic training. Alternatively, very recent works utilize deep reinforcement learning for restoring images with diverse or unknown corruptions. Though deep reinforcement learning can generate effective policy networks for operator selection or architecture search in image restoration, how it is connected to the classic deterministic training in solving inverse problems remains unclear. In this work, we propose a novel image denoising scheme via Residual Recovery using Reinforcement Learning, dubbed R3L. We show that R3L is equivalent to a deep recurrent neural network that is trained using a stochastic reward, in contrast to many popular denoisers using supervised learning with deterministic losses. To benchmark the effectiveness of reinforcement learning in R3L, we train a recurrent neural network with the same architecture for residual recovery  using the deterministic loss, thus to analyze how the two different training strategies affect the denoising performance. With such a unified benchmarking system, we demonstrate that the proposed R3L has better generalizability and robustness in image denoising when the estimated noise level varies, comparing to its counterparts using deterministic training, as well as various state-of-the-art image denoising algorithms. 
\end{abstract}
\section{Introduction}

Image denoising is one of the most fundamental inverse problems, which aims to estimate the underlying clean $\mathbf{x}$ from its noisy observation $\mathbf{y}$, which is corrupted with noise $\mathbf{n}$ as:
\begin{equation}
    \mathbf{y} = \mathbf{x} + \mathbf{n}\;.
\end{equation}
Assuming $\mathbf{n}$ to be the additive white Gaussian noise (AWGN), it follows the normal distribution, \ie,  $\mathbf{n} \sim \mathcal{N}(0,\,\sigma^{2})$. Besides improving the image visual quality, it is also a necessary preprocessing step for many high-level vision tasks such as classification~\cite{liu2018image}, segmentation \cite{liu2020connecting}, object detection~\cite{milyaev2017towards} and tracking~\cite{na2019mixture}. 


Classic image denoising algorithms are based on analytical models, \eg, image non-local similarity~\cite{10.1007/978-3-642-04146-4_9,dabov2007image,6909762}, transform-domain sparsity~\cite{liu2017image,wen2015structured}, etc.
More recently, deep learning has demonstrated remarkable results in image denoising by training the highly flexible neural networks with deterministic loss functions using an end-to-end approach~\cite{guo2019toward, Chen_2017,Zhang_2017,liu2018non}.
While most deep denoisers exploit feed-forward convolutional neural networks (CNNs)~\cite{Zhang_2017}, the models usually involve a huge amount of trainable parameters leading to high memory complexity.
Alternatively, some recent works exploit recurrent neural networks (RNNs) with shared module parameters. For example, the non-local recurrent network (NLRN) \cite{liu2018non} achieved both high parameter efficiency and denoising performance.

Comparing to the end-to-end supervised deep learning, few works to date exploited deep reinforcement learning (DRL) for image denoising.
Some pilot works trained a separate policy network for operator selection \cite{yu2018crafting,furuta2019fully} or architecture search \cite{yu2019pathrestore, vassilo2020multi} to assist image denoising.
However, it is unclear how DRL can be ``directly'' applied to inverse problems, \eg, how effective is the denoising network trained via DRL?
To the best of our knowledge, no work to date has benchmarked DRL with supervised deep learning with deterministic loss functions in image denoising.

To this end, we propose a novel image denoising scheme via Residual Recovery using Reinforcement Learning (R3L) for image denoising.
We show that the proposed R3L is equivalent to a RNN denoiser trained using a stochastic reward, which provides a unified framework to compare DRL to other RNN-based image denoising schemes.
To benchmark the effectiveness of DRL, we train a recurrent neural network with the same architecture as our R3L model for residual recovery using supervised learning with a deterministic mean square error, called R3N.
The experiments show that the proposed R3L achieved more reliable denoising results when the estimated noise levels (\ie, noise standard deviation $\sigma$) of degraded images deviate from the oracle. 
The average denoising PSNRs (over varied noise estimations) using our R3L outperform those by R3N as well as many state-of-the-art denoising algorithms. 

\section{Related Work}
Image denoising methods are classified into two categories: prior-based methods and learning based methods. Many classical methods, such as BM3D \cite{dabov2007image} and WNNM \cite{6909762}, are based on effective priors, and some of them applied the denoising operator recursively~\cite{6909762}.
On the other hand, learning-based methods utilized more flexible models such as deep neural networks~\cite{guo2019toward, Zhang_2017,zhang2018ffdnet}. Though deep denoising models lead to superior image restoration, most of them involve a huge amount of parameters. One solution to enhance memory efficiency is applying a lighter neural network recursively, which results in many successful frameworks based on RNN or DRL. We provide a summary of RNN and DRL algorithms for image denoising. 
Table \ref{tab:cap} summarizes the representatives of image denoising methods of different categories, as well as the proposed R3L.  
\subsection{RNN for Image Denoising}
Deep RNNs have been widely applied for image denoising. Chen~\etal~\cite{chen2016deep} first used a deep RNN which exploits temporal-spatial information for video denoising. Putzky~\etal~\cite{putzky2017recurrent} proposed a learning framework, dubbed Recurrent Inference Machines (RIM), in which they train a RNN to learn an inference algorithm for solving inverse problems. Liu~\etal~\cite{liu2018non} proposed NLRN which incorporates the non-local operations into an RNN for image restoration achieving the state-of-the-art results. 
However, most of the RNN-based denoising models are trained over a corpus of images containing the similar noise distribution, using the deterministic loss function (\eg, mean square error), thus hard to generalize to complex and inaccurately estimatesd noise in practice.
\subsection{DRL for Image Denoising}
DRL has recently gathered considerable interest showing great promise in many applications \cite{silver2016mastering}, including image processing tasks. Yu \etal \cite{yu2018crafting} firstly attempt to apply DRL to learn a policy to select suitable operators from a pre-defined toolbox to progressively restore corrupted images. 
Their improved version~\cite{yu2019pathrestore} can dynamically select an appropriate route for different image regions in a multi-path CNN, to perform spatial-varying image denoising. 
Furuta \etal \cite{furuta2019fully}  proposes pixelRL, the first framework to do a pixel-wise restoration. Most of DRL based methods still rely on manually designed filters. What the DRL agent learns is the order to apply filters instead of directly modifying the pixel values, \ie, residual recovery. Therefore, it remains unclear how DRL approaches to image denoising relate to other learning based methods.

\begin{table}
\centering
\caption{Comparison between various image denoisers, including the proposed R3L and other existing methods.}\label{tab:cap}
\begin{tabular}{|c|c|c|c|c|} 
\hline
\textbf{Methods} & \begin{tabular}[c]{@{}c@{}}Trainable \\kernels\end{tabular} & \begin{tabular}[c]{@{}c@{}}Residual \\learning\end{tabular} & Recursion & DRL  \\ 
\hline
BM3D~\cite{dabov2007image}    &   &             &     &     \\
\hline
WNNM~\cite{6909762}    &   &             & \checkmark     &     \\
\hline
DnCNN~\cite{Zhang_2017}   &    \checkmark   &   \checkmark  &     &     \\ 
\hline
NLRN~\cite{liu2018non}    &     \checkmark  &    \checkmark &  \checkmark    &     \\ 
\hline
pixelRL~\cite{furuta2019fully} &        &     &\checkmark    & \checkmark     \\ 
\hline
R3L     &   \checkmark          &       \checkmark     &  \checkmark   & \checkmark    \\ 
\hline

\end{tabular}
\end{table}

\section{Proposed R3L method}
\subsection{Residual Recovery as Markov Decision Process}
Residual recovery is commonly used in deep image denoising, which aims to obtain the residual image of the noisy input relative to the ground truth. As removing a residual can be considered  as sequentially removing several inter-residuals, residual recovery is a sequential decision problem. Therefore, we modeled the denoising problem via residual recovery as a Markov Decision Process (MDP), which can be solved using DRL. 


At each state $t$ ($t = 0$ denotes the initial state) of denoising, taking the noisy image ($t = 0$) or the denoised estimate ($t \geq 1$) from the previous state as the input $I^t \in \mathbb{R}^N$, the DRL agent follows a policy $\pi$ to output the probability $P(a_i^t|I^t)$ $\forall i$. 
Here $a_i^t$ denotes the estimated residual of the $i$th pixel ($1 \leq i \leq N$) at the state $t$. 
We apply a deep neural network to construct the policy $\pi$, denoted as the policy network with the trainable parameter $\theta_\pi$. 
$A$ is the action set, which consists of all discrete values in a predefined range, and $a_i^t \in A$. 
The estimated image is updated to $I^{t+1}$ by applying the output actions, and the agent can obtain a reward $r_i^t$ for each pixel. The denoising process repeats until the termination stage $n$, and outputs the final denoised image. The probability of an action trajectory $J_i$ for each pixel $i$, denoted as $P(J_i|I^0,\theta_\pi)$, is calculated as:
\begin{equation}
\begin{aligned}
   P(J_i|I^0,\theta_\pi) &= P(a_i^1|I^0,\theta_\pi)P(a_i^2|a_i^1,I^0,\theta_\pi)\\
    &...P(a_i^t|a_i^{t-1}...,a_i^1,I^0,\theta_\pi)\\
&=\prod_{t=1}^T P(a_i^t|J_i^{t-1},I^0,\theta_\pi)
\end{aligned}
\end{equation}
where $J_i \triangleq \left \{ a_i^1, a_i^2,\, ... \,, a_i^n \right \}$.

Following the common setting in DRL, we use the long-term discounted reward $R_i^t(J_i)$ to evaluate a policy at the state $i$, which is defined as: 
\begin{equation}
    R_i^t(J_i) \triangleq r_i^t+\gamma r_i^{t+1}+\gamma^2 r_i^{t+2}+ ... +\gamma^{n-t} r_i^{n}
\end{equation}
Here, $\gamma^j$ denotes the $j$th power of the discount factor $0 < \gamma < 1$, and $r_i^t$ denotes the reward for pixel $i$ at stage $t$.

The DRL agent can explore different trajectories towards learning the optimal policy $\pi^*$. Following $\pi^*$, the agent selects the optimal action at each state with the highest probability by maximizing the expectation of $R_i^0(J_i)$ as:
\begin{equation}
\begin{aligned}
     \pi^* = \underset{\pi}{\mathrm{argmax}}P(J_i|I^0,\theta_\pi)R_i^0(J_i)
\end{aligned}
\end{equation}
\subsection{Proposed R3L Framework}
Inspired by~\cite{furuta2019fully}, we apply the fully convolutional network (FCN) based asynchronous advantage actor-critic (A3C) \cite{mnih2016asynchronous} framework in the proposed residual recovery reinforcement learning (R3L) scheme. 
We apply FCN as the encoder which is widely used and effective in image processing tasks for the pixel-level modification. We apply A3C with a policy network $\pi$ and a value network $V$ to make the training more stable and efficient \cite{NIPS1999_464d828b}.

\begin{figure}[t]
\centering
\includegraphics[width=1.0\linewidth]{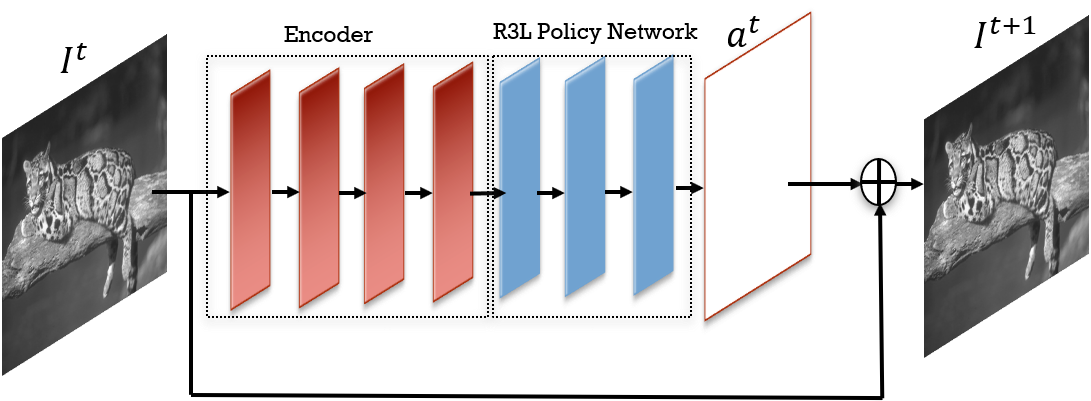}
\caption{The inference process of R3L at state $t$.}
\label{fig:inference}
\end{figure}

The FCN-based encoder is denoted as $\text{E}_\text{FCN}$, which is shared by both $\pi$ and $V$. 
$\text{E}_\text{FCN}$ extracts the features of the input image $I^t$ and outputs $s^t$, as the representation of state $t$.
Taking $s^t$, the policy network $\pi$ outputs the probability of selecting a certain residual value $a_i^t$ for each pixel, and the value network outputs $V(s^{t}|\theta_v)$, which is the estimation of the long term discounted rewards $R_i^t$ for each pixel. 
The reward $r_i^t$ used in R3L for image denoising is defined as:
\begin{equation}
    r_i^t \triangleq (x_i-I_i^{t-1})^2-(x_i-I_i^t)^2
\end{equation}
where $x$ denotes the clean image, and $x_i$ denotes its $i$ pixel.
Without loss of generality, for convenience, we consider the one-stage learning case ($n=1$) here. The gradients of the parameters of these two networks $\theta_\pi$,$\theta_v$ are calculated as:
\begin{equation}
\begin{aligned}
     R_i^t &= r_i^t+\gamma V(s^{t+1}|\theta_v)\\
     d\theta_v &= \nabla_{\theta_v}\frac{1}{N}\sum_{i=1}^N(R_i^t-V(s^t|\theta_v))^2\\
     d\theta_\pi &= -\nabla_{\theta_\pi}\frac{1}{N}\sum_{i=1}^N\log P(a_i^t|s^t,\theta_\pi)(R_i^t-V(s^t|\theta_v))
\end{aligned}
\end{equation}
During training, the residual value $a_i^t$ is sampled from a predefined range, \ie~ $\left[-13,13\right]$, according to the output from the policy network. 
In the testing phase, only the well-trained policy network is deployed and the residual value with the highest probability is greedily selected. The inference process is formulated as:
\begin{equation}
\begin{aligned}
&s^t = \text{E}_\text{FCN}(I^t)\\
&a^t = \text{Greedy}(\pi(s^t|\theta_\pi^*)) \quad \quad t =  0,1,2,...,T\\
&I^{t+1} = I^t+a^t
\end{aligned}\label{inference}
\end{equation}
where $\text{Greedy}(\cdot)$ denotes the deterministic greedy sampling operator \cite{kool2018attention,zhang2020deep}, $a^t$ denotes the residual image built by $a_i^t$ ($1 \leq i \leq N$), and $T$ denotes the number of total stages. Here, we use $T=5$ as a hyperparameter to balance the processing time and the performance. The inference process of R3L at state $t$ is illustrated as Fig. \ref{fig:inference}.
\subsection{Connection of R3L and RNNs}
\begin{theorem}\label{theorem1}
Greedily selecting the action with the highest policy from the output of policy network reduces the inference process of R3L to a RNN.
\end{theorem}
\begin{proof}\renewcommand{\qedsymbol}{}
In general, the recurrent inference process in an RNN is:
\begin{equation}\label{rnn_inf}
    I^{t+1} = f_\theta (I^t) \;\;\; \forall t\;,
\end{equation}
where $f_\theta$ is the recurrent module which is parameterized by $\theta$.
Based on (\ref{inference}), the inference process of R3L follows (\ref{rnn_inf}), with the corresponding module $f_\theta$ as the following form:
\begin{equation}
    f_\theta(I^t) = I^t + \text{Greedy}(\pi_\theta( \text{E}_\text{FCN}(I^t))).
\end{equation}
\end{proof}
Theorem \ref{theorem1} shows that the inference process of R3L follows an RNN. However, the R3L model is trained using DRL using the stochastic reward with no hidden states. In RNNs, the same network will be applied recursively until the termination. Our R3L also exploits the same recursive property and benefits a high parameter efficiency from it. However, most RNN based methods mainly focus on learning the final residual in an end-to-end manner, which makes the learning outcome a deterministic one-to-one mapping from the noisy input to the residual. R3L makes the solution a stochastic combination of different inter-residuals in a certain order rather than a deterministic mapping, and therefore has more flexibility. Moreover, RNNs use hidden states to summarize the modifications in the previous stages. In R3L, we assume that the action only depends on the current state input, so no hidden states are needed. 
\subsection{Benchmarking R3L with R3N}
Although R3L is connected to the existing learning based methods, since inference in the R3L is equivalent to applying an RNN, there is a lack of RNN based methods to do a fair comparison, because usually RNNs are combined with some other techniques and involve hidden states. To achieve a fair comparison and verify how the different training methods can help R3L, we propose a simplified RNN based benchmark named residual recovery RNN (R3N) for image denoising.

In the R3N, the input image $I^t$ is encoded via $\text{E}_\text{FCN}$ to $s^t$. Taking $s^t$ as input, a RNN block $\text{RNN}(\cdot|\theta_R)$ outputs the residual $\text{res}^t$, and $I^t$ is updated to $I^{t+1}$ by adding the residual to it. The whole process is formulated as:
\begin{equation}
\begin{aligned}
     &s^t = \text{E}_\text{FCN}(I^t)\\
     &\text{res}^t = \text{RNN}(s^t|\theta_R) \quad \quad t = 0,1,2,\dots,T\\
     &I^{t+1} = I^t + \text{res}^t\\
\end{aligned}
\end{equation}
and the gradient for the parameters of R3N is formulated as:
\begin{equation}
     d\theta_R = \nabla_{\theta_R}\frac{1}{N}\sum_{i=1}^N(I_i^{T+1}-x_i)^2
\end{equation}
where $T=5$ follows the same setting as R3L. \par
The inference process of R3N is basically the same as R3L shown in Fig.\ref{fig:inference}, but with the policy network replaced by the $\text{RNN}$ block. The specific design of the layers in R3N and R3L is summarized in Table \ref{tab:layer}. The numbers in the table denote the filter size, dilation factor, and output channels, respectively.
\begin{table}[htbp]
\centering
\caption{Specific design of the layers} \label{tab:layer}
\scalebox{0.72}{
\begin{tabular}{|c|c|c|c|c|c|} 
\hline
\multicolumn{2}{|c|}{$\text{E}_\text{FCN}(I^t)$}  & Conv+ReLU  & Conv+ReLU  & Conv+ReLU                                             & Conv+ReLU   \\ 
\cline{3-6}
\multicolumn{2}{|c|}{(in both R3L and R3N)}                          & 3x3, 1, 64 & 3x3, 2, 64 & 3x3, 3, 64                                            & 3x3, 4, 64  \\ 
\hline
\multirow{4}{*}{R3L} & \multirow{2}{*}{Policy~} & Conv+ReLU  & Conv+ReLU  & \multicolumn{2}{c|}{Conv+ReLU+Softmax}                              \\ 
\cline{3-6}
                     &                          & 3x3, 3, 64 & 3x3, 2, 64 & \multicolumn{2}{c|}{3x3, 1, \textbar{}A\textbar{}}                  \\ 
\cline{2-6}
                     & \multirow{2}{*}{Value}   & Conv+ReLU  & Conv+ReLU  & \multicolumn{2}{c|}{Conv}                                           \\ 
\cline{3-6}
                     &                          & 3x3, 3, 64 & 3x3, 2, 64 & \multicolumn{2}{c|}{3x3, 1, 1}                                      \\ 
\hline
\multicolumn{2}{|c|}{\multirow{2}{*}{R3N}}      & Conv+ReLU  & Conv+ReLU  &  \multicolumn{2}{c|}{Conv+tanh}    \\ 
\cline{3-6}
\multicolumn{2}{|c|}{}                          & 3x3, 3, 64 & 3x3 ,2, 64 & \multicolumn{2}{c|}{3x3, 1, 1}    \\
\hline
\end{tabular}}
\end{table}

\section{Experiments and Results}
\subsection{Experimental Settings}
We use the  BSD400 dataset as training images and BSD68~\cite{1467533} dataset as test images to verify the performance of R3L and R3N on image denoising. We train the models with additional Gaussian noise and the noise levels are selected as $\sigma = 25$ and $\sigma = 35$. However, in practice, it is difficult to estimate the noise level exactly accurately, and the noise level can vary in a range. Hence, besides testing the performance when the estimated noise level is accurate, we also test the cases when the noise level is estimated wrongly. For the model trained with noise level $\sigma = 25$, we test their performance when the noise levels are $\sigma$ = 15, 20, 25, 30 and 35. For the model trained with noise level $\sigma = 35$, we test their performance when the noise levels are $\sigma$ = 25, 30, 35, 40 and 45. Following the same setting, we also test the performance of BM3D, WNNM and DnCNN as baselines. The results are measured in terms of peak signal-to-noise ratio (PSNR) and shown in the next section.
\subsection{Results}

\begin{figure*}[ht]
\centering
\includegraphics[width=1.0\linewidth]{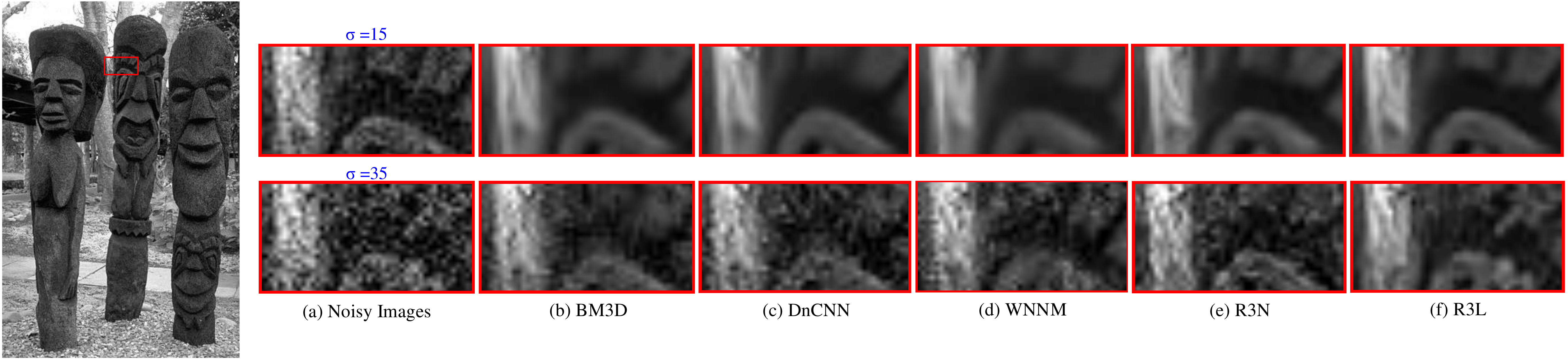}
\caption{Example of of denoising results using (b) BM3D~\cite{dabov2007image}, (c) DnCNN~\cite{Zhang_2017}, (d) WNNM~\cite{6909762}, (e) proposed R3N and (f) proposed R3L, with the zoom-in region highlighted. All the methods are set/trained with the estimated noise level $\sigma=25$. The first row are results for noisy images with $\sigma=15$ and the second row are results for noisy images with $\sigma=35$.}
\label{fig:sigma25d1535}
\end{figure*}

\begin{figure*}[ht]
\centering
\includegraphics[width=1.0\linewidth]{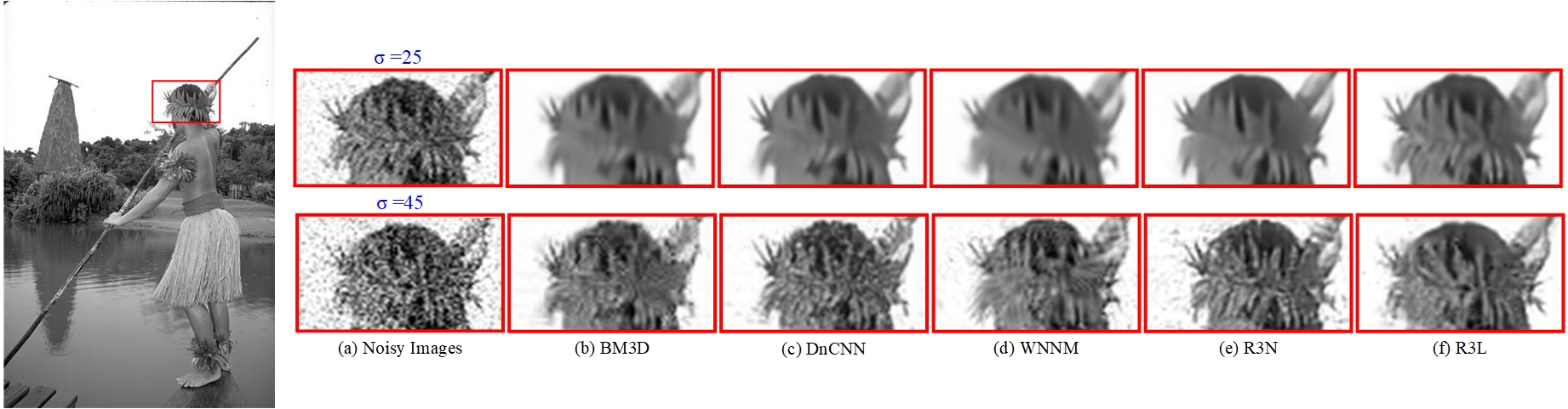}
\caption{Example of of denoising results using (b) BM3D~\cite{dabov2007image}, (c) DnCNN~\cite{Zhang_2017}, (d) WNNM~\cite{6909762}, (e) proposed R3N and (f) proposed R3L, with the zoom-in region highlighted. All the methods are set/trained with the estimated noise level $\sigma=35$. The first row are results for noisy images with $\sigma=25$ and the second row are results for noisy images with $\sigma=45$. }
\label{fig:sigma35d45}
\end{figure*}

\begin{table}[htbp]
\caption{The average PSNR (dB) results of different methods. All the methods are set/trained with $\sigma=25$. The best and the second best results are highlighted in \textcolor{red}{red} and \textcolor{blue}{blue} respectively. } \label{tab:25} 
\centering
\scalebox{0.86}{
\begin{tabular}{|c|c|c|c|c|c|} 
\hline
$\sigma$ & BM3D~\cite{dabov2007image}  & WNNM~\cite{6909762}  & DnCNN~\cite{Zhang_2017} 
& R3N            & R3L             \\ 
\hline
15  & 29.05 &  28.15    & \textcolor{blue}{29.17} & 28.83          & \textcolor{red}{29.64}  \\ 
\hline
20  & 28.87 &   28.57   & \textcolor{red}{29.42} & 29.07 & \textcolor{blue}{29.30}           \\ 
\hline
25  & 28.56 &    28.80  & \textcolor{red}{29.23} & \textcolor{blue}{28.95} & 28.73           \\ 
\hline
30  & \textcolor{red}{27.48}  &  26.94   & 26.40  & 26.70 & \textcolor{blue}{27.44}  \\ 
\hline
35  & \textcolor{blue}{24.88}   &   23.77 & 22.86  & 23.15 & \textcolor{red}{25.16}  \\
\hline
\emph{Average}&\textcolor{blue}{27.77} &    27.25 &  27.41  &  27.34  &\textcolor{red}{28.05}\\
\hline
\end{tabular}}
\end{table}
Table~\ref{tab:25} and Table~\ref{tab:35} summarize the PSNR results of our proposed frameworks and several
state-of-the-art denoising methods. It shows that though our proposed R3L does not perform the best when the estimation is accurate, it can outperform the baselines when the estimation is inaccurate. More specifically, for the cases when the estimation error is relatively large, for instance $\pm$ 10, R3L can still maintain a good denoising performance with a higher PSNR. It should be emphasised that R3L achieves these performance using far fewer parameters than DnCNN.\par Fig.\ref{fig:sigma25d1535} and Fig.\ref{fig:sigma35d45} demonstrate the visual quality of the denoised image for the different methods.
It shows that when the noise level is overestimated oversmoothing is a critical issue, however, the images processed by R3L have more detailed textures remaining. Moreover, when the noise level is underestimated, the denoised images from R3N and the other baselines still have obvious noise remaining and may also involve some artifacts, but R3L can remove most of the noise with no additional artifacts resulting in a more natural and better visual quality. 

The experimental results show that R3L is a more robust denoiser with high parameter efficiency. We explain the robustness from two points. First, compared with very deep end-to-end frameworks, R3L has less complexity, which endows R3L more generalization ability. Second, R3L is trained using a stochastic state-wise reward. The stochastic training process help R3L explore more different states and generate a more general policy than R3N, which is trained using a deterministic end-to-end loss. 

\begin{table}[htbp]
\caption{The average PSNR (dB) results of different methods.All the methods are set/trained with $\sigma=35$. The best and the second best results are highlighted in \textcolor{red}{red} and \textcolor{blue}{blue} respectively.} \label{tab:35}
\centering
\scalebox{0.86}{
\begin{tabular}{|c|c|c|c|c|c|} 
\hline
$\sigma$   & BM3D~\cite{dabov2007image}  & WNNM~\cite{6909762} & DnCNN~\cite{Zhang_2017}         & R3N            & R3L             \\ 
\hline
25 & 27.54  &  26.80    & \textcolor{blue}{27.68}          &  27.61          &\textcolor{red}{28.00}  \\ 
\hline
30 & 27.37  &  27.14    & \textcolor{red}{27.85}          &  27.74          & \textcolor{blue}{27.67}           \\ 
\hline
35 & 27.09  &  27.29    & \textcolor{red}{27.69}          &  \textcolor{blue}{27.44}          & 27.18           \\ 
\hline
40 & \textcolor{red}{26.32} &   26.08   & 25.68          &  25.58         & \textcolor{blue}{26.24}  \\ 
\hline
45 & \textcolor{blue}{24.39}  &   23.58   & 22.53          &  22.56         & \textcolor{red}{24.60}  \\
\hline
\emph{Average}& \textcolor{blue}{26.54}  &   26.18   &26.28                &  26.19          & \textcolor{red}{26.74}      \\
\hline
\end{tabular}}
\end{table}
\section{Conclusion}
In this paper, we propose a novel DRL based framework, namely R3L, to learn residual recovery for image denoising, and eventually close the gap of directly applying DRL for image denoising. We position R3L well by showing that R3L reduces to a deep RNN that is trained using the stochastic reward, and thus build the connection among R3L and the other methods. With the help of the proposed R3N, we benchmark R3L and verify how the different training method benefits R3L. The extensive experiment  results  reveals  that  R3L  is  a more  robust  denoiser  with  high parameter efficiency. Trained for a specific noise level, R3L can still be applied for a range of noise levels, which makes R3L a suitable framework for real-life scenarios, where the noise level estimation can be inaccurate.


\bibliographystyle{IEEEbib}
\bibliography{reference}

\end{document}